\documentclass[11pt]{article}
\usepackage{moriond,epsfig}

\def\Journal#1#2#3#4{{#1} {\bf #2}, #3 (#4)}

\def\AIP{\em AIP Conf. Proc.}

\def\be{\begin{equation}}
\def\ee{\end{equation}}
\def\bea{\begin{eqnarray}}
\def\eea{\end{eqnarray}}

\begin{document}
\vspace*{4cm}
\title{The INTEGRAL Science Data Centre}

\author{ V. Beckmann on behalf of the ISDC team }

\address{INTEGRAL Science Data Centre, Chemin d'Ecogia 16, CH-1290 Versoix, Switzerland}

\maketitle\abstracts{
The INTEGRAL Science Data Centre (ISDC) processes, archives and distributes data from the INTEGRAL mission. 
At the ISDC incoming data from the satellite are processed and searched for transient sources and Gamma-Ray bursts. The data are archived and distributed to the guest observers. As soon as the data are public, any astronomer can access the data via the internet. ISDC also provides the tools which are necessary for the data analysis and offers user support concerning questions related to the INTEGRAL data.
ISDC acts as a contacting point between the scientific community and the various instrument teams. In this proceeding an example for SPI data processing is shown.}

\section{Introduction}

Data from the INTEGRAL mission (see e.g. Winkler \& Hermsen~\cite{integral}), which is going to be launched October 2002, will be made available to the scientific community via the INTEGRAL Science Data Centre (ISDC; Courvoisier {\em et al.}~\cite{ISDC}). The ISDC is hosted by the Geneva observatory and is funded by an international consortium with ESA support. The ISDC is the contact point between the community, instrument teams, INTEGRAL Science Operations Centre (ISOC), Mission Operations Centre (MOC), the INTEGRAL project and the Russian data center. In this contribution we will discuss briefly the services provided by the ISDC, in terms of data and software which is made available, and show an example how the user can perform a scientific analysis using the INTEGRAL data with ISDC software. 

\section{INTEGRAL data flow through the ISDC}

\begin{figure}
\begin{center}
\psfig{figure=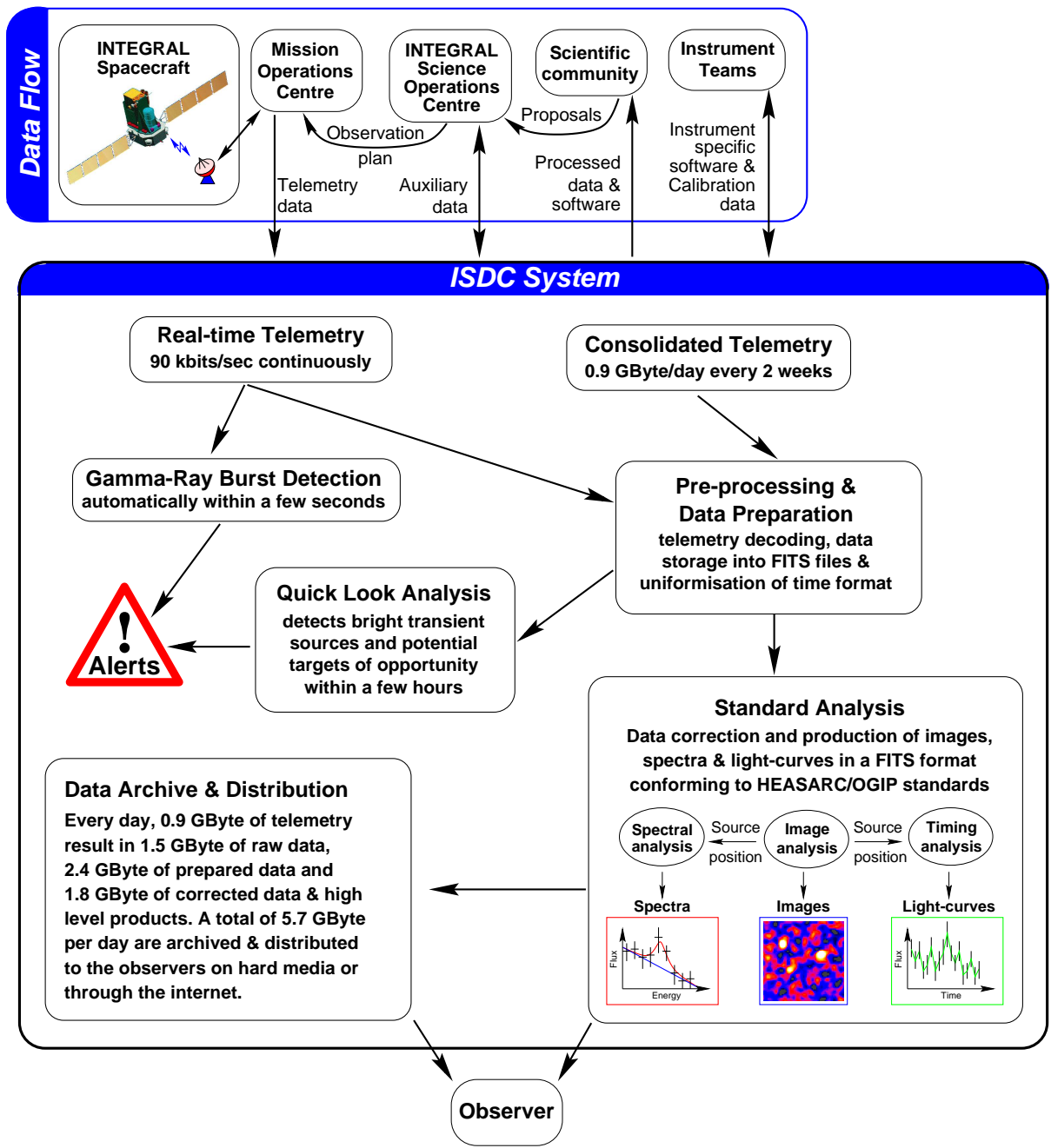,width=16.0cm}
\caption[]{\label{fig:dataflow}Data flow through the ISDC system. Graphic provided by Marc T\"urler (ISDC)}
\end{center}
\end{figure}
The INTEGRAL data flow through the ISDC is shown in Figure~\ref{fig:dataflow}.
The telemetry of the satellite is sent in a constant data flow of $\sim 90 \; \rm kbit/sec$ to the MOC. From there the data is then sent to the ISDC. At ISDC the data are processed in three ways:
\begin{itemize} 
\item The incoming data are searched for transient sources and Gamma-Ray bursts within a few seconds after arrival. This is done using the INTEGRAL Burst Alert System (IBAS; Mereghetti {\it et al.}~\cite{ibas}) will detect and localize gamma-ray bursts (GRB). Once a potential GRB is found, IBAS generates one or more GRB alerts. 
This functionality has already been successfully tested in a complete test run which involved INTEGRAL spacecraft and instruments, MOC, and the ISDC.
\item The telemetry data is then pre-processed, which means that it is decoded and data are stored into a fits file data structure. The time information is added to the events. These data are used for a quick look analysis. The purpose of this is to detect bright transient sources and potential targets of opportunity, and also gives a validation of the success of the performed observation. 
\item The processed data are also used to do a scientific standard analysis. This provides images and spectra in pre-defined energy ranges for the four instruments.
\end{itemize}

\section{ISDC service for guest observers}

The raw and processed data as well as the standard analysis output are made available to the principal investigator of the observation (e.g. the guest observer) as soon as possible. Data are distributed via the internet or by DVD/DLT. For the analysis of the data, ISDC also provides special software, which has been developed by the INTEGRAL instrument teams in collaboration with the ISDC. There are two main reasons for a special INTEGRAL software:
\begin{itemize}
\item The image deconvolution and source reconstruction for coded mask instruments like SPI (Jean {\em et al.}~\cite{SPI}), IBIS (Poulsen {\it et al.}~\cite{IBIS}), and JEM-X (Lund {\it et al.}~\cite{JEMX}) requires specific software (see e.g. Skinner, these proceedings). 
\item each observation is broken up into a set of single pointings of  $\sim 1000$ sec, the so-called dithering pattern. This is done to get a better background estimation and to achieve a higher spatial resolution for e.g. SPI. These single pointings then have to be combined afterwards, applying the pointing information as contained in the spacecraft housekeeping data.
\end{itemize} 
The different steps of scientific data analysis can be performed using the software provided by the ISDC.

\section{Hardware requirements and software installation}
The amount of hard-disk space needed depends on the length of the observation and is around 6 GByte per $10^5 \, \rm sec$ exposure time. Disk space for the data analysis is needed, which are additionally $\sim 2.5$ GByte per  $10^5 \, \rm sec$ of observation. 
The requirements on the hardware also depend on the type of analysis the user wants to perform. Especially the accessible memory (RAM) is important. For Solaris a work-station with $\sim 0.5$ GByte RAM should be sufficient, for Linux $\sim 1$ GByte RAM is recommended. 

The software can be downloaded via the internet\footnote{The ISDC web page is located at http://isdc.unige.ch/}. At the ISDC the pre-processing, quick-look, and standard analysis are performed under Solaris, but the software is also developed and tested as far as possible under Linux. 

The analysis of the data will make use of scripts which combine several steps and which make it possible for the user to re-run certain levels, as well as to display the output in a convenient way. These scripts are based on ROOT (see http://root.cern.ch/) which is developed at CERN. The whole procedure and technical details of the installation is described in a dedicated document, available via the internet from the ISDC web-page. 

After installing the software the data have to be copied from tape, DVD, or via the internet. 

\section{How to perform an analysis of INTEGRAL data}
In this section I will describe briefly how the analysis of INTEGRAL data will be performed. This should not be read as a user manual but rather as giving an idea what the INTEGRAL observer will have to face after receiving the data. 

The data provided to the observer can be used to perform a scientific analysis. The analysis scripts for the different instruments show the same structure and logic, even though they require different sets of parameters to be filled. I will show the scheme of the analysis for the spectrograph SPI, but in the main structure this is not different to the other INTEGRAL instruments. 

The single steps of the analysis can be run independently, calling single executable programs, or by running the script, which calls the single programs and fills some parameters (e.g. selecting the correct instrument calibration files).

The scheme of the SPI analysis (see also Strong, these proceedings) is shown in Table \ref{spidiagram}. The first column of this table gives the ISDC level of the analysis. These level names are the same for all instruments and allow the user to define start and stop levels when using the analysis scripts.
The first step is to apply the gain correction to the data, i.e. to determine the photon energy for each event which was detected by SPI. 
\begin{table}
\caption[]{Steps of the scientific analysis of INTEGRAL/SPI data}
\label{spidiagram}
\begin{tabular}{llll}
level & purpose & program\\
\hline
COR & gain correction of each detected event & spi\_gain\_corr \\
POIN & include spacecraft pointing information & spipoint\\
GTI & good time intervals & spi\_gti\_creation\\
DEAD & dead time correction & spidead\\
BIN\_I & defining the energy bins & spibounds\\
BIN\_I & event binning & spihist\\
BKG\_I & determine the background & spiback\\
IMA   & image deconvolution & spiskymax\\
IMA   & source reconstruction and light-curves & spiros\\
\end{tabular}
\end{table} 

Second step (POIN) is to extract the pointing information as it is stored in the spacecraft housekeeping data. This level creates a fits table with an entry for each dithering pointing including the spacecraft attitude and the start and stop time.

The next level (GTI) is the determination of the good time intervals (GTI) of the instrument. The GTI is e.g. influenced by telemetry gaps, solar flares, or could also be user-defined (e.g. excluding a certain time of the observation from the analysis). The output is stored in a fits table which lists the GTI per pointing. 

Level four (DEAD) determines the dead time for the given GTI. The dead time is based on the fact that the electronics of the instrument is busy for a fraction of the observation time and during this time no events are detected. Furthermore the dead time is depending on the rate of false events (e.g. photons which are not coming through the mask but through the side shield of the instrument), which are detected by the SPI Anti-Coincidence Shield (ACS). The information about the dead time fraction and the total life time for each detector is also stored in a fits table. 

In the BIN\_I step the user defines the energy bins in which the analysis should be performed. For imaging this might be a rather wide binning, while for receiving a spectrum this could result in several hundreds of energy bins. These energy bins are used in the further analysis. First the events are binned according to the user's selection and detector spectra (shadow-grams) are produced.

This is followed by the determination of the background of the instrument during the observation (BKG\_I). Especially SPI and IBIS are background dominated, which means that the correct subtraction of the background is essential for receiving meaningful results. The background could either be fed into the analysis process by applying a model, or by making use of several measured values, as the count-rate from the ACS or from the detector count-rates itself. 

At this level now all information is available to perform an image deconvolution or source reconstruction. 
One tool to perform the image deconvolution is {\em spiskymax} (see Strong, these proceedings) which is applying a maximum entropy method to the SPI data. 
To search for sources and to reconstruct source spectra, the user can apply {\em spiros} (developed by Connell, University of Birmingham) which is using the Iterative Removal of Sources (IROS) method. The spectra then can be used with XSPEC for spectral model fitting. {\em spiros} also offers the possibility to produce light-curves of sources.  

For both programs the user can apply a catalogue with known sources to the analysis, or search for new sources using the IROS method.

\section{Status of the analysis software}
The software has been tested on calibration and simulated data. Nevertheless scientific validation on in-flight data is necessary in order to provide software which gives the user confidence in the analysis outputs. Therefore the first software release to the scientific community will take place several months after launch (a working package will be made available to the teams involved in the commissioning phase activities and in the verification phase earlier). 

In any case the principal investigator receives the results of the standard analysis shortly after the observation has been performed. 


\section*{Acknowledgments}
We would like to remind that the work presented here is a team effort achieved by the INTEGRAL instrument teams and the ISDC.



\section*{References}
\begin{thebibliography}{99}
\bibitem{integral}C. Winkler and W. Hermsen, \Journal{\AIP}{510}{676}{2000}
\bibitem{ISDC} T. Courvoisier, {\it et al.}, {\it Proc. 3rd INTEGRAL Workshop - Astro. Lett. and Communications}, {\bf 39}, 355 (1999).
\bibitem{ibas}S. Mereghetti, D.I. Cremonesi, J. Borkowski, {\em Proc. 4th INTEGRAL Workshop}, eds. A. Gim\'enez, V. Reglero \& C. Winkler, ESA SP-{\bf 459}, 513 (2001)
\bibitem{SPI}P. Jean, G. Vedrenne, V. Sch\"onfelder, {\it et al.}, {\it Proc. 5th Compton Symposium}, {\bf 510}, 708 (2000)
\bibitem{IBIS}J.M. Poulsen, P. Sarra, A.J. Bird, {\it et al.}, {\em Proc. 4th INTEGRAL Workshop}, eds. A. Gim\'enez, V. Reglero \& C. Winkler, ESA SP-{\bf 459}, 615 (2001)
\bibitem{JEMX}N. Lund, N.J. Westergaard, S. Brand, {\it et al.}, {\it et al.}, {\it Proc. 5th Compton Symposium}, {\bf 510}, 727 (2000)
\end{thebibliography}
\end{document}